# A comparison of low-cost behavioral observation software applications for handheld computers and recommendations for use

Running title: Behavioral observation software application review

Annemarie van der Marel, Claire L. O'Connell, Sanjay Prasher, Chelsea Carminito, Xavier Francis, Elizabeth A. Hobson

Department of Biological Sciences, University of Cincinnati, Cincinnati, OH, 45221, USA

Orcid ID:

van der Marel: 0000-0003-3942-8314

O'Connell: 0000-0002-3852-3021

Prasher: 0000-0001-6175-5747

Carminito: 0000-0002-0432-3865

Francis: 0000-0001-7286-3284

Hobson: 0000-0003-1523-6967

*Corresponding author*: A. van der Marel, Department of Biological Sciences, University of Cincinnati, Cincinnati, OH, 45221, USA. Email: avdmarel@outlook.com

## Abstract

1. In the field of animal behavior and behavioral ecology, many standardized methods to observe animal behavior were established in the last decades. While the protocols remained similar, behavioral researchers can take advantage of technological advancements to enter observations directly onto a handheld computer (phone, tablet, etc.), saving time and potentially increasing fidelity of recordings. However, we now have the choice between many different platforms for recording behavioral observations. Our challenge is choosing the most appropriate platform that fits a particular study question, research design, budget, and desired amount of preparatory time.

2. Here, we review six low-cost software applications for handheld computers that are available for real-time entry of behavioral observations: Animal Behaviour Pro, Animal Observer, BORIS, CyberTracker, Prim8, and ZooMonitor.

3. We discuss the preliminary decisions that have to be made about the study design, and we assess the six applications by providing the advantages and disadvantages of each platform and an overall application comparison. In our supplemental material we illustrate the setup and data collection routines, and how to customize certain platforms so they will work more effectively for particular study aims or sampling methods.

4. Our goal is to help researchers make calculated decisions about what behavioral observation platform is best for their study system and question.

## Introduction

In the field of animal behavior and behavioral ecology, a very important aspect of collecting data is direct behavioral observations. Since the foundational work of Altmann (1974), clear methodologies exist for the direct collection of observations (classic behavioral and ecological monitoring methods) and many researchers follow these methods (17440 Google Scholar citations, accessed 22-10-2021). However, the past five decades have seen major advances in technology that have changed the ways in which researchers collect and analyze data. Prior to the widespread use of computers, researchers used pen and paper to record behavioral observations and perform analyses. Today, the availability of computers and especially handheld computers in the field allow for rapid real-time data collection, which can be entered directly into electronic format and is virtually ready for analyses.

The more recent technological advancement of automated methods, such as GPS trackers and proximity loggers, provide an exciting avenue to collect behavioral and ecological data (Shizuka et al., 2020; Smith & Pinter-Wollman, 2020). Nevertheless, traditional observation methods are necessary to validate these automated methods (Gelardi et al., 2020) and to describe and collect data on the contexts of behaviors (something most automated methods still cannot collect).

Multiple software applications (apps) have been developed to collect behavioral observations. In this review, we focused on low-cost (<$100.00USD) mobile applications that allow for real-time entry of behavioral observations onto a handheld computer (Animal Behaviour Pro, Animal Observer, BORIS, CyberTracker, Prim8, and ZooMonitor). Other applications, such as video coding software programs or more costly programs, are mentioned in Supplemental Material 1. The reviewed software platforms differ in how the behaviors are entered, what sampling methods are supported, what contextual variables can be entered, the output for analysis, the speed of data entry, and the operation system used (requirement of Android or Apple products). We conducted an initial comparison of these apps for our own work on monk parakeet social structure (van der Marel, Prasher, et al., 2020) and found that the decision about which app to use for a particular study question is often complicated. To aid others in making similar decisions, we compiled this review to compare the features and benefits of each of these apps. We provide descriptions of: 1) how to make initial decisions about the study design, 2) the

advantages and disadvantages of each of these six apps, and 3) an overall app comparison, which will allow researchers to make calculated decisions about which app would work best for their study question, sampling methods, and budget.

## Preliminary decisions about study design

Many initial decisions are critical to evaluate during the study planning phase. The research question will influence which sampling method will be used and thus the app choice. Characteristics of the study site may influence how often the data can be uploaded to the cloud or whether back-up material, such as external hard drives, are necessary. When starting with a new study system, researchers may have to develop an informative ethogram from preliminary observations or customize ethograms of closely related species to better represent their study system. All the applications we discuss record observational data but differ in how they record contextual data or comments. Therefore, researchers should decide what information is necessary: only behavioral data (e.g., actor, behavior (either states or events), subject, and time) or whether they also need contextual data (e.g., habitat characteristics, weather, predominant group activity, group size and membership). How behavior is entered can affect the speed of data entry, which may be crucial for accurate data collection or a personal preference. Another thing to keep in mind is the data output. Many researchers prefer to import their data into the statistical program R (R Core Team, 2020). Output that is immediately ready for import to R, such as comma-delimited (CSV) or text files, would eliminate any potential errors made during intermediate data entry steps.

Another major point to consider is the sampling method, which depends on the research question (Altmann, 1974). The primary sampling method is scan sampling, where behaviors are sampled from the whole group at preselected time points. Scan sampling is very useful to get the percentage of time individuals perform a specific behavior. All-occurrence/behavioral sampling (only specified behaviors are recorded whenever they occur) is especially useful for behaviors that are of interest but do not occur often, such as social behaviors for certain species (Altmann, 1974). Researchers can combine both scan and all-occurrence sampling to get both information on a daily activity budget and social interactions. Sometimes it is better to record as much as the observer can (i.e., ad libitum sampling), although biases in sampling may occur as observers may unconsciously record more conspicuous behaviors or may observe certain group members more readily. Focal sampling requires an observer to focus on one individual or one group for a specified amount of time. Instead of focusing on a focal individual, the observer could also focus on a sequence of interactions. When the direction and asymmetry of interactions are of interest, observers could record the interaction data between dyads (i.e., sociometric matrix completion). Research questions often require more and/or fine scale data, thus the method in which the data is collected is crucial for efficiently and accurately collecting data. Most behavioral observation recording applications have the option to perform scan, all-occurrence, ad libitum, and focal animal sampling. However, some of the applications only provide an opportunity to perform scan and focal animal sampling.

Finally, the features available within an app, the app's compatibility with a particular device, and the budget also warrants consideration. Some applications are only compatible with iOS, and others with Android or Windows desktop, and some are free while others are not. Below, we provide a summary of the aforementioned considerations for the low-cost behavioral observation software applications for handheld computers, Animal Behaviour Pro, Animal Observer, BORIS, CyberTracker, Prim8 and ZooMonitor (see Table 1).

## Assessment of the applications

Here we discuss the advantages and drawbacks of using each application and compare them regarding the duration of research projects, the app's learning curve, sampling methods and personal preferences. To assess the usage of the different apps, we looked up the number of times the apps were cited (either by searching for the app name or the papers introducing the app) in Google Scholar.

**Table 1**. Overview and comparison of the application software.

| Application[a] | Sampling methods | Data input type | Entry methods | Output file format | Customization | Operation system (app version) |
|---|---|---|---|---|---|---|
| **Animal Behaviour Pro**[a] | Focal animal, all-occurrence, ad libitum, scan | Behavioral | Codes as buttons on screen | CSV files[c] | N/A | iOS (v1.5) |
| **Animal Observer** | Focal animal, all-occurrence, ad libitum, scan[b] | Behavioral, contextual, GPS | Swiping + drop-down | dat files → CSV files | Multiple layout options for observation session | iOS (v1.0) |
| **BORIS** | Focal animal, all-occurrence, ad libitum, scan | Behavioral, contextual | Codes as buttons on screen | In-app analyses, multiple options, incl CSV[c] | Nearly unlimited | Multiplatform desktop app (v7.10.5), Android |
| **CyberTracker** | Focal animal, ad libitum, scan[b] | Behavioral, contextual, GPS | Customizable | Multiple options, incl CSV[c] | Nearly unlimited | Android (v 1.0.377; iOS via CyberTracker Connect) |
| **Prim8** | Focal animal, all-occurrence, ad libitum, scan | Behavioral, contextual | Codes on keypad | CSV files[c] | Separate keypad | Android (v1.0) |
| **ZooMonitor**[a] | Focal animal, all-occurrence, ad libitum, scan | Behavioral, contextual | Codes as buttons on screen + map | In-app analyses, CSV files[c] | N/A | Web-based able to use on any platform (v3.3.9) |

[a] All apps are open-source except for Animal Behaviour Pro (0.99US$) and ZooMonitor for non-accredited institutions (50US$/year). [b] multiple methods simultaneously, [c] readily available for R

### Animal Behaviour Pro

Animal Behaviour Pro was developed by Nicholas Newton-Fisher and first released in 2012 (Newton-Fisher, 2020, https://apps.apple.com/us/app/animal-behaviour-pro/id579588319). The app can accommodate a wide range of behaviors and has been used across taxa including birds (Xie et al., 2017), humans (Dunbar et al., 2017), mammals (Roberts et al., 2016), and primates (Berthier & Semple, 2018; Boeving et al., 2020; Tórrez-Herrera et al., 2020). A search for "Animal Behaviour Pro app" resulted in 10 citations on Google Scholar (accessed October 22, 2021).

### Advantages and benefits of Animal Behaviour Pro

Animal Behaviour Pro allows for a quick data input (users can input specific behaviors with modifiers), highly customizable coding schemes, multiple sampling methods, and multiple export options (Table 1). A unique four-digit login code identifies the observer recording the behaviors, which is useful when researchers need to discriminate between data collected by different observers. Another advantage is that behaviors can be categorized to help organization and ease of data entry. For example, multiple coding schemes can be created and stored for different projects or sampling periods. Within a particular coding scheme, a user can enter an unlimited number of subjects and codes, which appear as buttons on the screen. Observers can choose their own codes of any length (2-4 characters are recommended for visibility, but this is not required). Furthermore, behaviors can be classified as a "state" or an "event", and the user can specify whether the behaviors are mutually exclusive. Additionally, users may add color tags to the type of behavior or specific subjects for easy recognition. The app also includes the option to add "modifiers" to any behavior for a more detailed description of any behavior. Furthermore, the flexibility of the coding scheme allows for quickly recording coalition behavior. When creating a coding scheme, observers can add a button to include two individuals that make up a coalition so during sampling periods both individuals can be entered simultaneously. These features make the app highly configurable, allowing it to be adaptable to various study systems and research questions.

### Potential drawbacks or other considerations of Animal Behaviour Pro

Although Animal Behaviour Pro is highly customizable, some drawbacks exist. Animal Behaviour Pro allows for many specifications, and although useful, it can overcomplicate the initial set up of coding scheme(s) particularly when creating the scheme within the app. A YouTube tutorial ("Getting Started with Animal Behaviour Pro" https://www.youtube.com/watch?v=v9pLE9kpOnk) is available that breaks down the basics of the features available which is convenient for overcoming the initial learning curve;

we also provide a walkthrough for the setup for focal animal sampling in Supplemental Material 2.1. Another potential drawback is the detailed coding schemes with many behaviors or individuals that could cause researchers to spend much time scrolling for a specific behavior particularly when the screen is small (phone instead of tablet). This process may slow down the data entry and behaviors may be missed as one has to look down at the screen for longer periods of time. Additionally, coding a single button to represent multiple individuals means those individuals will be accounted for in a single cell in the data output. In this case, the data would need to be reformatted which could be time consuming. Alternatively, users may group many subjects together and sort and cluster the buttons for actors and receivers (start group names with 1, 2, 3, etc.). Furthermore, sampling does not start until the observer enters an observation which eliminates dead time at the beginning of a sampling period. However, this could be a potential drawback for researchers interested in comparing the timing of the first event and the beginning of the sampling period. Another drawback is that users cannot edit behaviors once submitted. Unless specified in the "Notes" (which corresponds with a single observation), the app does not provide a way to document contextual data such as weather or quality of observation period. This app does not provide any location specific features, however, the virtually limitless coding scheme allows observers to enter a specific location as a subject and record which individuals are present at that location, or use codes to record who is, for example, resting near or walking with whom. However, as mentioned above, this method could require adding several new codes and potentially take time away from observations. No analyses can be done within the app. Finally, this app produces a CSV file in which the output of the scan and ad libitum sampling includes all subjects with their behavior in the same row (Supplemental Material 3), a data format which can complicate further analyses. Finally, the software is only available under iOS and thus requires Apple hardware.

### Animal Observer

Animal Observer was designed by Damien Caillaud and his team from the Dian Fossey Gorilla Fund International and first released in 2012. The application has been used to collect behavioral data in, for example, primates (Harrison et al., 2020; Schrock et al., 2019), humans (Dai et al., 2020), bats (Welch et al., 2020), and birds (van der Marel, Prasher, et al., 2020). A search for "Animal Observer" Caillaud resulted in 12 citations on Google Scholar (accessed October 22, 2021).

### Advantages and benefits of Animal Observer

The Animal Observer app has a very user-friendly interface, which allows accurate data collection of the location of individuals. This application is useful for study systems where many interactions can occur in a short period of time, and when individual spatial (GPS) locations are required. The Animal Observer app is initialized with the Animal Observer Toolbox (https://fosseyfund.github.io/AOToolBox/sidebar-left.html). This toolbox is a web application that uses R in the background, which allows researchers to customize Animal Observer. It may take a while to optimize the exact needs but there are many options to include different variables, and it is very customizable. For example, in species where pairs are the primary social unit, pairs can be added as "IDs" which allows for entering behaviors that occur between pairs instead of individuals. Thus, researchers face a trade-off between the potentially time-consuming initial set up and optimizing the app to best suit their research question (Supplemental Material 2.2).

Another advantage is the daily data collection setup as it provides a visual representation to collect behavioral data. In a scan, the user can move the ID from the left field to the map/navigation field, after which the ID button can be clicked upon which a drop-down menu of behaviors pops up. During focal sampling, the observer can move between IDs on the map and automatically a drop-down menu appears with behavioral data. Also, users can correct any of the variables if a mistake is made during a focal sample. Any additional information (notes) can be easily added using the built-in voice recorder, camera, or text editor. This app has two options to export the data, either using a cable or a sftp (Secure File Transfer Protocol) server. Care must be taken to delete data from the Ipad/Ipod once downloaded, otherwise duplicate entries in the data files will appear. The sftp server allows for the behavioral data to be uploaded automatically to a server without having to connect the device to a computer to copy the data files. The implementation of these types of functions increases data reproducibility. The observation CSV file is immediately useable in R as each row represents one observation (Supplemental Material 3). Be aware that if any of the variables are changed throughout the field season, the order of the variables in the output CSV files may end up in different columns, making it more difficult to collate the files from different observation sessions into a single file for analyses.

### Potential drawbacks or other considerations of Animal Observer

A potential drawback is that this application was developed for studying primates, and in the primate literature, focal animal sampling is the main sampling method. However, in other taxa, scan and all-occurrence sampling are used more frequently. Thus, for studies where interaction data is the focus, Animal Observer may not be the first choice. However, users can customize the application to focus on all-occurrence sampling and collect scan sampling at the same time (for a detailed description of how to set up both sampling methods, see Supplemental Material 2.2). Also, at the end of each scan, observers should not forget to hit 'End scan' after the behaviors of all individuals in sight are entered, otherwise the scan behaviors will not be saved. Animal Observer does not have an option to perform in-app analyses, so all the data need to be downloaded and used in other programs. A drawback of this process is the amount of time it takes to import and export the datafiles (the created files need to be converted from CSV to json (JavaScript Object Notation) so it can be used in the app on the iPad and the behavioral data files from json to CSV). Finally, the software is only available under iOS and thus requires Apple hardware.

## BORIS (Behavioral Observation Research Interactive Software)

BORIS software was developed by Olivier Friard and Marco Gamba and first released in 2016 as a multiplatform desktop app (Friard & Gamba, 2016). The desktop version has been cited 982 times on Google Scholar (accessed on October 22, 2021). An Android app version is current under development and available for use and testing. The desktop version of BORIS has been used in diverse animal taxa, e.g., insects (Taylor et al., 2021), fish (Wing et al., 2021), and birds (Prasher et al., 2019) and is typically used to score videos or direct observations of animals in captivity. Here, we review the in-development version of the Android app which accommodates the recording of live observations from hand-held devices.

## Advantages and benefits of BORIS

BORIS provides many features that make it a flexible tool for recording a wide range of behavioral data (Table 1). Furthermore, as it is an open-source software with the source code available on GitHub, the level of customization is potentially limitless. The BORIS Android app is designed to be used with the BORIS desktop software, which is available across platforms (Windows, MacOS, and Linux; http://www.boris.unito.it/pages/download.html). The user interface is easy to use, and user guides are available on the BORIS website to help researchers understand how the program works (see also Supplemental Material 2.3 for a walkthrough). Another advantage is the daily data collection setup, where behaviors can be set as point or state events and each behavior can be associated with two or more modifiers for an additional level of specificity. Observers can also specify which behaviors are mutually exclusive. If groups of animals are being observed within the same observation period, a list of subjects can be added to the BORIS project. Subjects can also be defined as pairs of individuals or groups to record behaviors occurring above the individual level. Behaviors, modifiers, and subjects defined in the ethogram will appear as separate boxes on screen when using the Android app. If behaviors are assigned into different behavioral categories (e.g., attack and preen may be categorized into a 'social' category), they will appear in boxes with a distinct color for each category. Finally, contextual data (e.g., weather, group size, observer ID, etc.) that researchers want to record at the beginning of every observation can be specified at this stage. Once an ethogram is made, it can be exported in various formats to share with other observers. Alternatively, the whole project file itself can be shared with other observers so that everyone has the same ethogram, list of subjects, and contextual variables. After an observation session, particular events can be edited (e.g., changing the behavior or adjusting the time) by exporting the observation to the BORIS desktop. Comments can be added to specific events within the BORIS desktop, but there is no built-in way to take notes on things that fall outside of what was defined in the BORIS project (e.g., behaviors not in ethogram – although a behavior called "other" could be added to the ethogram to record any unusual activity).

Another advantage is that some in-app analyses are possible after transferring completed observations to the BORIS desktop. For example, researchers can access time budgets for each observation, which include data such as the number and duration of events per subject and inter-event intervals. If applicable, time budgets for behavioral categories defined in the ethogram will also be available. Time budgets can be easily exported as plain text files or spreadsheets. Additionally, coded events can be plotted by time within BORIS for a quick visualization of the data. Frequency and duration of behaviors can be plotted as well, and plots can be exported in various formats. Finally, behavioral data can be exported in a variety of formats (including CSV) for further analysis in other programs.

Potential drawbacks or other considerations of BORIS

The main drawback of BORIS is that the software is primarily available as desktop software. The Android app is still at an early stage of development and there is no version for other mobile platforms (e.g., iOS). Additionally, as the mobile app is still in development, some of BORIS' useful features (e.g., using behavioral coding maps, adding comments within observations or editing events) are not yet available. Full use of the app with the BORIS desktop requires an internet connection to send data to and from the Android and desktop apps. Finally, because of the number of features, it may take some time to get accustomed to using the program, especially when making more detailed observations.

## CyberTracker

The CyberTracker software was designed by CyberTracker Conservation Organisation and initially released in 1997 (available at https://www.cybertracker.org/software/free-download for all Microsoft Windows versions). The app was not primarily designed to collect behavioral observation data but instead to collect GPS field data (Liebenberg et al., 2017). However, CyberTracker is used for behavioral observations across many animal taxa, e.g. lizards (Carter et al., 2012), birds (Ashton et al., 2019), primates (Burgunder et al., 2017; Castles et al., 2014; Herzog et al., 2014; Marshall et al., 2015; Martina et al., 2020) and other mammals (Marneweck et al., 2015; Rauber et al., 2019; Venter et al., 2019; Welch et al., 2018). A search for "CyberTracker animal behavior" resulted in 606 citations on Google Scholar (accessed October 21, 2021).

### Advantages and benefits of CyberTracker

CyberTracker is highly customizable for many types of research projects, especially for collecting GPS data (Table 1 and see Supplemental Material 2.4 for a walkthrough). The CyberTracker software functions as a tool to create and design a mobile application which allows researchers to customize an app interface to best suit the researchers' needs. Users design the application using the free desktop software. The desktop software provides a test run option so researchers can review their design before finalizing the app interface and export it to the mobile device (Android). Observers can edit an entry during the sampling period by going to the main screen and selecting 'all sightings' where the wrong entry can be edited. Collected data can be returned to the computer for extraction and analysis (Supplemental Material 3 for an example of the data output). CyberTracker was designed to accommodate an inclusive user base: app setup requires little to no coding skills, and icon-based user interfaces to allow for participation of non-literate users. CyberTracker is useful in several fields of ecology and biology due to its versatility. This app was developed in 1997 and is still actively updated and supported, therefore this app may be useful when planning for a long-term research project.

### Potential drawbacks or other considerations of CyberTracker

Despite its versatility, the highly customizable nature of the app can be overwhelming particularly at first encounter, and it may require a substantial time commitment to properly set up the application interface for your needs. To remedy this the CyberTracker official website has extensive resources available for reference, including step-by-step instructions for setting up a basic app interface (we provided a walkthrough in Supplemental Material 2.4). As mentioned previously, the CyberTracker desktop software is only compatible with Android devices. iOS functionality is currently in development (https://archive.cybertracker.org/software/getting-started).

## Prim8

Prim8 mobile was developed by Monica McDonald and Scott Johnson and initially released in 2014 (McDonald & Johnson, 2014, http://www.prim8software.com). It was primarily developed to observe primate behavior (McDonald & Johnson, 2014; Whitehouse et al., 2017; Whitehouse & Meunier, 2020). However, the app has also been used for studies with other mammals (Szott et al., 2019; van der Marel et al., 2019, 2021; van der Marel, Waterman, et al., 2020) and birds (Benti et al., 2019; Puehringer-Sturmayr et al., 2020; Szipl et al., 2019). A search for the paper introducing the app resulted in 25 citations on Google Scholar (accessed October 21, 2021).

### Advantages and benefits of Prim8

Prim8 allows for great flexibility in the coding scheme as users can make-up code for anything they want as Prim8 is code-based, which is more similar to the old-fashioned pen and paper method. For example, users can collect data on pairs of individuals or groups instead of individuals, depending on how the actor and recipients are coded. Also, every observer could make up their own code (as long as the names of the individuals and behaviors are the same across devices) because the codes are immediately translated to the names once an observation has been entered in the data entry field. Furthermore, observers can add many different variables as the number of modifiers after the fixed position of actor-behavior-(recipient) is unlimited. This flexibility also allows an observer to enter any other comments at the end of the fixed code positions. Thus, Prim8 is a good choice when users are comfortable with entering data on a qwerty keyboard and do not want to spend a lot of time customizing the app.

Another advantage is that the setup and daily behavioral data collection are very straightforward (see Supplemental Material 2.5 for a walkthrough). An observer can use the application immediately from an Android device, without having to connect it to a computer. Also, observers are prompted to answer whether the behavior (1) has a recipient (if so, then the mandatory code field reflects "actor-behavior-recipient" and if not, it reflects "actor-behavior" after which any modifiers can be added), (2) is a state vs an event, (3) is an all-occurrence behavior, (4) or a scan behavior. Users can get the duration of state behaviors if they click on the behavior once the state has finished. If a mistake is made, observers could click on the behavior instance in question (all behavior instances are visible below the data entry field after hitting enter) and retype the behavior. An observer can input notes by simply starting the note using an exclamation mark. For example, a note can be entered as "! hawk circles above". Users can connect their Android device to their computer and transfer the files. The output files (nine different CSV files that include the different sampling methods and backup files) can be directly imported in any statistical software program reducing intermediate data entry steps (see Supplemental Material 3).

### Potential drawbacks or other considerations of Prim8

Although this app is customizable in how to record behaviors, some drawbacks exist. First, users must remember the codes that were set for the behaviors and individuals. Thus, the training of multiple observers may take time. However, the codes are provided in the individuals and ethogram tabs, so that users can look up codes during an observation session. Another potential drawback is the usage of modifiers. If the same modifiers are not entered in the same sequence after the mandatory code, the modifiers may end up in different columns once the data is exported, which users will have to adjust manually later. Furthermore, the keypad provides a flexible way to record behaviors. But keypads are not always responsive or easy to use. The developers of Prim8 suggest using a phone with a physical keypad instead of a screen, but this phone may not be readily available. Alternatively, observers could use their own Android smartphone or tablet and purchase an external wireless Bluetooth keyboard. Prim8 does not have any in-app analyses. Finally, although the website states that the app is also freely available on iOS, as of October 2021 the app is currently only freely available on Android platforms.

### ZooMonitor

ZooMonitor was developed by Lincoln Park Zoo and was initially released in 2016 (Ross et al., 2016; Wark et al., 2019, https://zoomonitor.org) to collect data that may impact animal welfare and inform management decisions (Ross et al., 2016; Wark et al., 2019). This app is often the preferred data collection platform for many Association of Zoos and Aquariums (AZA) accredited zoos, aquariums, and sanctuaries. The paper by Wark et al. (2019) introducing the app has been cited 20 times, on Google Scholar (accessed October 22, 2021).

### Advantages and benefits of ZooMonitor

The ZooMonitor application is highly customizable, allowing users to group behaviors, add modifiers to behaviors, upload a grided map, and use multiple sampling methods at once. These attributes allow for the app to be especially useful for tackling research questions based around the space use, activity budget, or social relationships of both a small and large number of focal individuals.

Another advantage is the relatively low learning curve, allowing for many researchers to be trained quickly (see Supplemental Material 2.6 for a walkthrough). Within the setup steps of the ZooMonitor Admin section, users can specify the details of the study (name of study, study length, focal animals, etc.) as well as the used sampling methods. These sampling methods are treated as separate "channels" within ZooMonitor and can all be employed simultaneously. For each sampling method channel, users can specify the respective behaviors that fall under that specific sampling method (e.g., aggression might be a behavior that a researcher wants to record every instance of, so it would be entered under both the all-occurrence channel and the scan sampling channel). Users could add up to three modifiers per entry in the behavior details pop up window, in which it is also possible to add social modifiers and a description of the behavior in question. If desired, users can upload a map image under the Space Use channel. The user can then overlay a grid atop the map with a customizable number of grid columns and rows to best divide the map into sections that may help with space use research questions. After setup completion of all the necessary channels, users may now create additional questions that will be answered before each observation session to specify contextual data, such as weather, temperature, and crowd size. Users can score both group and individual behavior during observation in the ZooMonitor App section. ZooMonitor also features a separate module to record ad libitum observations. Ad libitum observations may be recorded as open text, numbers, or via a user defined list of options that was predetermined by the user in the initial setup phase. After data collection is complete, in-app analysis tools allow users to begin the data analysis process within the app itself.

Potential drawbacks or other considerations of ZooMonitor

Even though ZooMonitor is highly versatile and can be used to tackle a wide variety of research questions, it does have limitations that make its use in certain situations not ideal. First, while ZooMonitor's streamlined nature allows for ease in mastery of the app, it also limits the possibilities a user might have when analysing data in-app. ZooMonitor's built in analysis tools are rigid and are not customizable, only meant for a preliminary look at the data via graphs before in-depth analysis, and thus may not be useful for the observer's research question. In this case the in-app analysis tools would have to be ignored and the data can be exported for analysis elsewhere. In addition, once data has been collected, it cannot be edited to fix potential mistakes. A user must instead make a note of the error and edit the data once it has been exported. Finally, ZooMonitor costs significantly more than other software applications unless the users are part of an AZA accredited institution. For some, the app's higher price point may overshadow its potential uses.

## Application comparison

The choice of app best suited to a particular study question and a particular research team depends on multiple aspects. Considering the duration of the study, the learning curve, which sampling methods are used, the output, but also personal preferences, can affect which app is the most suitable for each study (Table 1). First, the duration of the research project may influence the app choice. For a short-term study, such as a short-term undergraduate summer project, researchers may want to consider an app with a shorter learning curve, such as Animal Behaviour Pro, BORIS, Prim8, or ZooMonitor. However, for a long-term research project, researchers could invest more time in an app that will be supported over a longer period and is still actively developed and updated, such as CyberTracker, but which has a greater starting cost.

A second consideration is the learning curve to set up and use the application. Sometimes prior knowledge is required to use an app, such as running code in R for Animal Observer. Other apps allow users to use customized coding schemes, so that observers that are not fluent in a certain language can use their own codes (Animal Behaviour Pro and Prim8). CyberTracker accommodates an inclusive user base: the icon-based user interface allows for participation of non-literate users. Although CyberTracker's setup may be straightforward when used for the collection of ecological and GPS data, the setup is more difficult in the context of behavioral data collection. Of course, observers also need to be trained using the application and this training may be faster when the behaviors are presented as buttons on a screen (Animal Behaviour Pro, BORIS, CyberTracker, or ZooMonitor) or as a drop-down menu (Animal Observer) compared to codes entered on a keyboard (Prim8).

Other factors to keep in mind is what sampling methods will be used, the number of animals that will be observed, what information will be entered, personal preferences, the data output and budget (Table 1). For example, researchers may prefer Android over iOS devices or vice versa, or code-based over point-and-click data entry methods. All the apps' output files can be exported as CSV files, but the output is very different across platforms (Supplemental Material 3), which may influence the app choice.  A major advantage of all the reviewed behavioral observation entry software applications is that they are low-cost.

## Conclusions

In this contribution, we compared six low-cost behavioral observation software applications for handheld computers, which eliminate transcription time and can reduce potential errors made while transcribing the data. We emphasized the advantages and benefits as well as the methodological concerns involved in picking the most appropriate app for a particular study design. This approach highlights where each application excels as well as where study designs or other considerations may lead researchers to prefer one app over another. Our overview and the guidelines we provide will be useful for researchers who are evaluating which behavioral observation software application may work best for their study and will make that decision easier and less time-consuming.

## Acknowledgement
We thank the application developers of BORIS and ZooMonitor, and an anonymous reviewer for their feedback. We thank Science Twitter for their advice on applications for behavioral data collection.

### Authors' contributions

AM and EH designed the research; AM, CO, SP, CC, XF reviewed the different applications and performed sampling; AM led the writing of the manuscript. All authors contributed critically to the drafts and gave final approval for publication.

### Data availability

In our supplemental material, we provided information about the setup and the output files of each app using focal animal sampling of horses of the Wild Discoveries program at the University of Florida (https://programs.ifas.ufl.edu/wild-discoveries/research-practice-materials/ accessed January 22, 2021). We provided the setup files (ethogram and individual files) and the output for each application in a Github repository (https://github.com/annemarievdmarel/behavioral-observation-data-entry-apps). We expect that this information will be helpful to researchers who are weighing decisions about which app to choose and will make that process easier and less time-consuming.

### Conflict of interest

We declare that we have no conflict of interest as we decided to include the software applications without receiving any funds from these companies or software developers.

Dai, S., Bouchet, H., Nardy, A., Fleury, E., Chevrot, J. P., & Karsai, M. (2020). Temporal social network reconstruction using wireless proximity sensors: model selection and consequences. *EPJ Data Science*, *9*(1). https://doi.org/10.1140/epjds/s13688-020-00237-8

Dunbar, R. I. M., Launay, J., Wlodarski, R., Robertson, C., Pearce, E., Carney, J., & MacCarron, P. (2017). Functional benefits of (modest) alcohol consumption. *Adaptive Human Behavior and Physiology*, *3*(2), 118–133. https://doi.org/10.1007/s40750-016-0058-4

Friard, O., & Gamba, M. (2016). BORIS: a free, versatile open-source event-logging software for video/audio coding and live observations. *Methods in Ecology and Evolution*, *7*(11), 1325–1330. https://doi.org/10.1111/2041-210X.12584

Gelardi, V., Godard, J., Paleressompoulle, D., Claidiere, N., & Barrat, A. (2020). Measuring social networks in primates: wearable sensors versus direct observations. *Proceedings of the Royal Society A: Mathematical, Physical and Engineering Sciences*, *476*(2236), 20190737. https://doi.org/10.1098/rspa.2019.0737

Harrison, N. J., Hill, R. A., Alexander, C., Marsh, C. D., Nowak, M. G., Abdullah, A., Slater, H. D., & Korstjens, A. H. (2020). Sleeping trees and sleep-related behaviours of the siamang (*Symphalangus syndactylus*) in a tropical lowland rainforest, Sumatra, Indonesia. *Primates*, *0123456789*. https://doi.org/10.1007/s10329-020-00849-8

Herzog, N. M., Parker, C. H., Keefe, E. R., Coxworth, J., Barrett, A., & Hawkes, K. (2014). Fire and home range expansion: A behavioral response to burning among savanna dwelling vervet monkeys (*Chlorocebus aethiops*). *American Journal of Physical Anthropology*, *154*(4), 554–560. https://doi.org/10.1002/ajpa.22550

Liebenberg, L., Steventon, J., Brahman, !Nate, Benadie, K., Minye, J., Langwane, H. (Karoha), & Xhukwe, Q. (Uase). (2017). Smartphone Icon User Interface design for non-literate trackers and its implications for an inclusive citizen science. *Biological Conservation*, *208*, 155–162. https://doi.org/10.1016/j.biocon.2016.04.033

Marneweck, D., Cameron, E. Z., Ganswindt, A., & Dalerum, F. (2015). Behavioural and endocrine correlates to the aardwolf mating system. *Mammalian Biology*, *80*(1), 31–38. https://doi.org/10.1016/j.mambio.2014.08.001

Marshall, H. H., Carter, A. J., Ashford, A., Rowcliffe, J. M., & Cowlishaw, G. (2015). Social effects on foraging behavior and success depend on local environmental conditions. *Ecology and Evolution*, *5*(2), 475–492. https://doi.org/10.1002/ece3.1377

Martina, C., Cowlishaw, G., & Carter, A. J. (2020). Exploring individual variation in associative learning abilities through an operant conditioning task in wild baboons. *PLOS ONE*, *15*(4), e0230810. https://doi.org/10.1371/journal.pone.0230810

McDonald, M., & Johnson, S. (2014). "There's an app for that": a new program for the collection of behavioural field data. *Animal Behaviour*, *95*, 81–87. https://doi.org/10.1016/j.anbehav.2014.06.009

Newton-Fisher, N. E. (2020). *Animal Behaviour Pro* (Version 1.5; p. Mobile app).

Prasher, S., Evans, J. C., Thompson, M. J., & Morand-Ferron, J. (2019). Characterizing innovators: Ecological and individual predictors of problem-solving performance. *PLOS ONE*, *14*(6), e0217464. https://doi.org/10.1371/journal.pone.0217464
21